\documentclass[10pt]{article}
\usepackage[top=1.0in, bottom=1.0in, left=0.75in, right=0.75in]{geometry}

\title{A Survey of Virtualization Techniques\\
Focusing on Secure On-Demand Cluster Computing}
\author{Nadir Kiyanclar \\
University of Illinois at Urbana-Champaign \\
National Center for Supercomputing Applications \\
\{\textit{nadir@ncsa.uiuc.edu}\}}

\begin{document}
\maketitle

\abstract{Virtualization, a technique once used to multiplex the
resources of high-priced mainframe hardware, is seeing a
resurgence in applicability with the increasing computing power of
commodity computers.  By inserting a layer of software between the
machine and traditional operating systems, this technology allows
access to a shared 
computing medium in a manner that is secure, resource-controlled, and
efficient.  These properties are attractive in the field of on-demand
computing, where the fine-grained subdivision of resources provided by
virtualized systems allows potentially higher utilization of computing
resources.

It this work, we survey a number of virtual machine systems with the
goal of finding an appropriate candidate to serve as the basis for the
On-Demand Secure Cluster Computing project at the National Center for
Supercomputing Applications.  Contenders are reviewed on a number of
desirable properties including portability and security.   We conclude
with a comparison and justification of our choice.}

\pagebreak

\section{Introduction}
Virtualization\footnote{this is virtualization in the traditional
sense.  The systems surveyed in this paper actually follow a number of
different strategies in providing a virtual execution environment to
applications} is the faithful reproduction of an entire architecture
in software which provides the illusion of a real machine to all
software running above it.  A virtual machine monitor (VMM) manages
the creation, destruction and control of one or more virtual machines
(VM) on a computer.  The VMM is responsible for controlling access to
the resources of the real hardware, as well as multiplexing the
execution of multiple VMs fairly. As defined in~\cite{smith04virtual},
the term 
virtualization is widely applicable, and is also used to describe a
number of other technologies used for portability and compatibility in
a large number of applications.  `Virtual Machines' are noted in the
contexts as diverse as those of the Java platform and VMWare for
example.  

Virtualization is the realization of a common theme in
computer science: adding a layer of indirection between the user of a
computer and computer hardware in the name of flexibility.  The key
innovation in a virtualized system is the choice of where to perform
the indirection: between the hardware and the software which
originally ran on it.  Probably the most well-known example of this
principle is seen in VMWare, a system which virtualizes the Intel
IA-32 architecture, thus allowing users to run multiple copies of
commodity operating systems such as Windows and Linux hosted on a
so-called `native' OS\footnote{This brief description refers to the
hosted versions of VMWare; ESX server uses a different architecture
and is described in detail later in this paper.}  The traditional
definition of virtualization as laid out by
Goldberg~\cite{goldberg73architecture} 
suggests that a virtual machine monitor differs from emulation in that
the implementation must not only be complete in its simulation of a
given machine architecture, but also efficient.  If a virtual machine
monitor recreates the same architecture as that of the physical
machine it runs on, this can entail running all VM code on the native
hardware to as great a degree as possible, and trapping all
instructions which read/write privileged machine state to maintain
isolation between VMs.  Otherwise, binary translation must be used to
maintain efficient execution of the VM's code.
Some of the key concepts behind virtualization - namely, encapsulating
the state of a running VM to provide efficient isolation, resource
access, and execution - while useful in their own right, also hold
great potential in other applications requiring efficient resource
management.  In particular, virtualization holds great promise for the
field of on-demand Computing.

%
%

\subsection{Enabling On-Demand Computing}
The term ``Grid Computing'' originated in the 1990s, and gained
prominence with the publication of 
the seminal works on the subject, "Physiology of the Grid" and
"Anatomy of the Grid"~\cite{foster01anatomy,foster02physiology}.  The
motivation for the Grid is as a medium over which scientists could cooperate to solve
difficult problems in physics, mathematics, and medicine, to name just
a few applications. The most well-known implementation of the principles of
Grid computing is the Globus Toolkit~\cite{globusweb}, developed and
provided by the Globus project.  The Globus Toolkit embodies a `bag of
tools' approach to developing Grid applications: software is provided
to manage security (GSS-API), secure file transfer (GridFTP), and
resource allocation and resource management (GRAM) among others.  Wide
deployment of Globus and similar systems has driven development in
related areas of collaborative scientific computing, one such area
being \textit{On-Demand Computing}.  As discussed
in~\cite{koenig04design}, on-demand computing attempts to address the
problem of fluctuating demands for computing resources.  Acting
individually, organizations must either allocate sufficient resources
to deal with peaks in user demand (and thus accept severe
under-utilization of resources at non-peak times as a consequence), or
accept resource limitations at peak demand times.  An on-demand computing
infrastructure would let organizations with differing demand cycles
cooperate by spilling and accepting load as appropriate to alleviate
resource shortages: When organization $A$, for example, is
experiencing a peak, organization $B$, whose resources are
underutilized, can accept some of $A$'s load in exchange for
compensation of some sort.  

These issues are being addressed as part of the On-Demand Secure Cluster
Computing (ODSCC) project at the National Center for Supercomputing
Applications.  The overall motivation and 
framework for this project is laid out in~\cite{koenig04design}.
Currently this system is implemented on top of the
Faucets~\cite{kale04faucets} system, with additional facilities for load
balancing across clusters~\cite{koenig05using}.  We are interested in
extending this work using virtualization to enhance its generality and
security.  In the balance of this section, we discuss
some of the advantages to be had by adding a layer of abstraction to a
an on-demand computing system via virtualization.  Note that on-demand
computing systems can be viewed as constructs on top of some of the
facilities provided by a Grid infrastructure, and therefore we
sometimes may use the terms `Grid computing' and `On-demand computing'
interchangeably below, although the terms are not strictly equivalent.

\subsection{Advantages of Virtualization}
Applications must specifically be enabled to take advantage of
on-demand computing systems, a result of the fact that the unit of
abstraction of a system resource is inherently an application-level
construct.  In today's computing environment, this is is clearly a
requirement, since modern operating systems do not support the
features required for an on-demand infrastructure.  Among these
features are checkpointing, migration, an accounting system for
tracking resource usage between organizations, and security
restriction of privileged OS code.  The application-level nature of
grid facilities leads to inevitable incompatibilities between
competing implementations of on-demand computing systems.  The end
result is that creating an application which can take advantage of a
on-demand computing infrastructure can require porting legacy code among
different systems in order to take advantage of the features of each.
Moving code from one system to another requires similar
efforts.~\cite{huang03adaptive} describes a system which can provide
near source code-level compatibility for legacy MPI applications being
migrating to run on the Charm system\footnote{thus these ported legacy
applications can take advantages of the checkpointing, migration, and
load balancing features in Charm.  Porting is still required, however}
Virtualization can remove this burden more generally from grid
application developers by moving essential On-Demand computing-enabling
mechanisms below the layer of the executing distributed application.
This notion was first proposed in the context of Grid computing
in~\cite{figuieredo03case}, and many of the issues discussed below are
shared and further developed in that work.  To highlight:

\begin{itemize}
\item{Transparency}: Assuming the necessary controls are implemented
in the Virtual Machine Monitor, features such as process migration and
checkpointing could be seamlessly added to a distributed program with
no requirement on the part of the application developer.

\item{Legacy Support}: By implementing a subset of the required
functionality at the VMM layer, legacy distributed applications could
be supported on a On-Demand computing substrate, providing these
applications with the load balancing and and efficiency benefits such
a system has to offer. Optional functionality could be deferred to
application-level libraries used by programs running inside a VM, and
could be used to optionally improve the performance of distributed
applications.  The benefit to developers is a smoother migration path
for legacy code. 

\item{Simplicity}: The separation of mechanism and policy is a
well-tested principle in software engineering.  Recent work in virtual
machine research and resource kernels takes this approach to an
extreme, controlling only access to machine resources and leaving all
policy decisions to applications, which in the case of VMMs are
usually full OS instances.  In the Exokernel paper, these application
level OSes are referred to as LibOSes~\cite{engler95exokernel}.  In
terms of code complexity, the benefits of decoupling abstraction and
resource control, as done in research kernels and VMMs, can be
striking: While the code base of modern full-fledged operating systems
can be comprised of millions of lines of code, VMMs such as
Disco(bugnion97disco,govil00cellular) and Xen are comprised of tens of
thousands of lines of code, a considerable reduction in
complexity\footnote{both of these operating systems take advantage of
the device drivers of native operating systems, so this figure does
not represent lines of driver code, only the abstract interfaces
provided to virtual machines}.  Applying this strategy to a VM-based
On-Demand system may have similar benefits.  By implementing low-level
`Resource allocation primitives' in such a system below the layer of
the VM, the implementation complexity of resource allocation functions
for distributed applications can be greatly reduced in upper layers.
As an example, the developers of the uDenali system were able to
implement a simple VM migration mechanism in 289 lines of C code using
the functionality already present in that
system~\cite{whitaker05rethinking}.

\item{Monitoring}: Observing the state of the executing components of
a distributed computation is potentially a more reliable process in
the context of virtual machines. The ability to observe program state
from `outside' an active VM would let developers monitor the state of
virtual machines even in the case of an application crash, as a
connection can be maintained to a trusted monitoring system at the VMM
level in such a case.  \cite{ho04pdb} describes a system for VMM-based
monitoring on a single machine, but the concept could be extended to
facilitate distributed debugging.

\item{Security}: Related to the above point, monitoring below the VM
level allows secure auditing and logging of VM state and actions
without VM knowledge, in a strongly secure manner\footnote{it may
still be technically possible to exploit a flaw in the VMM to crash or
compromise the VMM logging system in such a system.  However, the
small size of the VMM mentioned above makes the security auditing of
VMM code a much more manageable process than, for example, that of
auditing the Linux kernel source.}  Similarly, other security controls
and access restrictions can be implemented in the VMM layer with
less fear of a compromise of the security system itself.  The
SHype~\cite{sailer05shype} features such a design, which is used to
provide secure resource access and delegation, and the VAX VMM
security kernel~\cite{seiden90auditing} uses a VMM-monitoring approach
to provide a secure logging and auditing mechanism.
\end{itemize}

\section{Evaluation Criteria}
A useful strategy in implementing a VM-based Grid resource management
system is to build on existing virtual machine monitor
implementations, and to extend such software to work in the context of
a Grid.  In order to do so, existing VM Monitor software must be
evaluated with regards to the applicability and desirability of its
use in such a system.  To that end, evaluation criteria must be
determined.  The above points, along with the performance requirements
which partially drive the development of collaborative Grid systems,
lead to the following points for looking at the VM systems in this
paper:

\begin{itemize}
\item{Performance}: Clearly, high performance is a necessity in a Grid
system, as much of the drive towards computational Grids in the first
place was influenced by a need for more computing cycles.  Certainly
the definition of an acceptable reduction in computing performance
varies with the organization running a computation, and with the
application itself.  In general, however, care should be taken that
the indirection provided by the use of virtualization does not come
at too high a cost.

\item{Trust}: The reliability and security hinted at earlier make the
assumption that the virtual machine monitor is itself reliable and
immune to attack.  While it was argued above that a `traditional' VMM
can be much smaller in the size of its code base than an OS, and thus
much easier to audit, not all of the systems herein use a full-system
virtualization strategy.  While more the minimal OS and library-level
virtualization techniques these systems employ can have a benefit in
terms of performance, they are potentially harder to verify as being
free of security-related bugs.

\item{Portability}: Part of the motivation given above for a VM-based
Grid system was the potential ease with which legacy code could be
accommodated.  With full-system virtualization, this legacy support is
trivially achieved; the VMM presents an interface to application
software which is identical to that of a real machine; thus an
existing OS need not be modified at all to run on such a VMM.  On the
other hand, process-level virtual machines require interception at the
OS level, and paravirtualized architectures such as
Denali~\cite{whitaker02scale} and Xen~\cite{dragovic03xen} require an
OS to be ported to their respective architectures. 

\item{Multiplicity}: A VM-based Grid system reduces the complexity of
starting jobs on physical machines because multiple jobs can run
concurrently on the same physical machine.  The effectiveness of such
a strategy depends on the support available in a VMM for executing
multiple VM instances.  This support consists of two components: the
efficiency with which this sharing is done, and the isolation provided
between instances of VMs.
\end{itemize}

The balance of this paper is devoted to a more in-depth exploration of
the criteria above.  Examples from virtualization systems developed
from the 1960s through the present are used to illustrate the
different motivations for using system virtualization and the
techniques used to overcome obstacles to the implementation of virtual
machine monitors on various architectures.  After this, all of the
systems reviewed are compared side-by-side.  The primary motivation
for this on-demand-computing oriented survey is to aid in choosing a
virtualization system on top of which to implement a VM-based
on-demand on-demand computing infrastructure.  In the conclusion we
briefly outline our design, elucidate the reasoning behind our choice
of VMM software to build on, and discuss the resulting considerations.

\section{Survey}
\subsection{Performance}
In~\cite{popek74formal}, Goldberg explicitly defines efficiency as a
requirement for a virtual machine monitor to fulfill.  The method
behind this efficiency is also mentioned: use the native hardware of the
physical machine to as great a degree as possible.  A goal for any
system chosen as the foundation of a VM-based on-demand computing
architecture is that virtualized performance be as close to native
speed as possible.  It should be noted that some of the techniques
described in this section, when coupled with on-line optimizations, can
actually increase the speed of code running under virtualized
execution.  HP's Dynamo~\cite{bala00dynamo} is an example of such a system,
under which even statically optimized code could see up to a 20%
performance boost.  Dynamic optimization is outside the scope of this
paper, however, so we will use the word `efficient' in the sense of
near-native execution speed.  A number of factors involving both
software and hardware impact the performance of virtualized systems
negatively, and depending on hardware features and compatibility
requirements, various techniques are used to combat these issues.

Under a fully virtualizable architecture, no modifications are
necessary to run a `native hardware' OS under virtualization.  The IBM
System/370~\cite{seawright79study} is an example of a system
architecture which is not only 
virtualizable, but which was designed (as an extension to the older
System/360) with virtualization in mind~\cite{creasy81origin}.
Unfortunately for modern-day 
VMM researchers, most modern ISAs were not designed with
virtualization in mind, and consequently are not fully virtualizable
according to the definition laid out by Popek and Goldberg.  A
simple summary of these requirements is that privileged instructions
(those which modify or read privileged system state) behave differently
depending on the privilege level at which they are executed.  In
supervisor mode, they should behave as normal, but execution in user
(or problem) mode should cause a software trap to a control program.
This behavior allows a VMM to run OS software in user mode along with
applications, and automatically receive control whenever the OS
attempts to read or set privileged system state.  The VMM can then
perform the operation on behalf of the guest OS, and manipulate the
guest's state to maintain the illusion that the guest is in control of
real hardware.

As stated above, most architectures (including the IA-32) define
instructions which do not fulfill these requirements for
virtualizibility.  On the IA-32 for example, certain instructions which
read and write privileged state in supervisor mode act as no-ops in
user mode, or perform a restricted (but trap-free) operation instead.
On these architectures, special measures must be taken to run a truly
isolated full-system virtual machine.  Certain VM systems use binary
translation to deal with non-virtualizable instructions.
VMWare~\cite{vmwareweb} is the most well-known example of such a system,
and comes in a variety of versions for applications ranging from
workstation to enterprise (ESX Server).  While the features and
implementations of each vary, the general principles of executing
code are the same: code segments are scanned and instructions
identified as non-virtualizable are either replaced with jumps into
the VMM or with sequences of equivalent instructions which trap or
safely perform the original operation.  Three factors mitigate the
speed overhead of binary translation: First, the translation done is
relatively simple, as it involves substituting selected instructions.
Second, translated segments are stored in a code cache.  Third, only
what are termed `sensitive segments' of code need be translated; the
rest can be directly executed on the hardware.  `Sensitive segments'
are usually synonymous with operating system code running inside a VM;
direct execution does not change the behavior of user programs, which
run in user processor mode both under a VMM and on native hardware.
In addition to CPU virtualization, I/O devices also require isolation
in a VM.  Interrupt-driven I/O can incur a large expense in
virtualized systems, because most application I/O requests incur a
trap not only into the guest OS, but also eventually into the the VMM
to perform the `real' I/O.  VMWare Workstation's hosted architecture
incurs an even larger performance overhead due to the additional
mapping of all VM I/O requests to host kernel requests, as detailed
in~\cite{venkitachalam01virtualizing}. 

An alternate strategy for dealing with a non-virtualizable ISA is to
modify the source code of an operating system to work cooperatively
with a VMM running in privileged processor mode.  The operating system
would effectively be using a new ISA, similar to but distinct from the
original, which replaces all non-virtualizable instructions with calls
directly into a VMM.  While OS modification is required if this method
used, the similarity of the implicitly defined
`virtualization-friendly' ISA makes this porting (in theory) a
manageable exercise.  The preceding description refers to what is
called paravirtualization in the literature.  Paravirtualization
sacrifices some portability in the name of efficiency, and is used in
a number of systems, explicitly or otherwise.

The Xen virtual machine monitor~\cite{dragovic03xen} interacts with
its guest OS 
instances via a paravirtualized architecture very similar to the x86
ISA.  Xen,
as opposed to Denali below, is targeted at working with existing
operating systems.  Consequently the paravirtualized architecture in
that system shares many more of the features of real hardware than
Denali.  Explicitly porting each guest OS to Xen has a significant
performance benefit, as optimizations such as interrupt batching can
reduces CPU overheads from that source.  The feasibility of porting an
OS to a paravirtualized architecture is confirmed
in~\cite{dragovic03xen}, where it is
noted that a small group of researchers was able to manage a port of
the Linux 2.4 kernel to Xen\footnote{further proof is offered by the
existence of stable or development ports to Xen of version 2.6 of the Linux
kernel, FreeBSD, NetBSD, and Plan 9}.  Performance benefits can be
achieved by taking advantage of the fact that a paravirtualized OS is
`aware' that it is running in a VM.  For example,
drivers can be created for guests which map to explicitly virtual
devices.  These devices can be designed to result in a simple
implementation and efficient operation, avoiding some some of the
overheads of virtualizing a `real' device, as noted
in~\cite{venkitachalam01virtualizing} .  Note
that VMWare ESX server offers a similar
option~\cite{waldespurger02memory}.  Finally, the I/O
multiplexing performed by Xen transfers data in page sized units using
zero-copy techniques.  The end result is that in benchmarks and more
general testing, a Linux 2.4 kernel running under Xen can achieve within 5\% of
the performance of a natively running kernel.  In~\cite{menon05xen},
 it is shown
that when using very high-throughput devices (such as gigabit ethernet
adapters) to back virtual devices, VM I/O can become CPU-bound at high
transfer rates, artificially bounding throughput in a manner similar
to that observed in VMWare Workstation
in~\cite{venkitachalam01virtualizing} .  However, this phenomenon
happens at a significantly higher peak transfer rate than in VMWare
Workstation. 

Denali~\cite{whitaker02scale} also uses paravirtualization, but the
ends and implementation reflect differing goals from that of the Xen
VMM.  While authors of Xen test up to 16 instances of virtual machines
on the Xen VMM, the goal for Denali is to scale to thousands of VM
instances.  The virtual ISA, while similar to x86, originally had some
significant departures from that architecture, such as the lack of a
VMM-provided `virtual MMU;' the original intention was to run a large
number of lightweight `libOS' applications, in a manner inspired by
the Exokernel~\cite{engler95exokernel}.  More recently such a virtual
MMU has been added, and with it the ability to support traditional
operating systems such as NetBSD and Linux (a NetBSD to Denali
currently port exists).  However, the implementation of virtual memory
under Denali greatly differs from either Xen's paravirtualized ISA or
IA-32; the interface presented to a VM is that of a software managed
TLB, rather than the hardware assisted paging unit of the
IA-32\footnote{such architected TLBs are common in RISC architectures
such as the MIPS or DEC Alpha}.  Published tests for Denali focus on a
customized libOS called Ilwacos.  Therefore, a direct comparison to a
native IA-32 OS in terms of raw performance is difficult to make.
Discussion of the scalability gains achieved with Denali is deferred
to section~\ref{perf-multiplicity}  Table~\ref{perf_comparison_table}
briefly summarizes the results of this section.
%
%
~\\
\begin{table}
\caption{Performance Summary}
\begin{center}
\begin{tabular}{| p{2in} | p{2in} |}\hline\label{perf_comparison_table}
\textbf{System}	& \textbf{Performance vs. Native}\\
\hline
VMWare ESX Server& n/a\\
\hline
Denali & n/a\\
\hline
Xen    & 95\%$+$ of native\\
\hline
VServer & Identical\\
\hline
User-Mode Linux & 10$-$100\% of native performance, depending heavily
on degree of system calls in code\\
\hline
Disco & 90\%+ of native\\
\hline
Cellular Disco & 90\%+ of native\\
\hline
\end{tabular}
\end{center}
\end{table}

\subsection{Portability}
Any on-demand computing system which strives to be the basis of a global
computing infrastructure must be widely deployable, or risk relegation
to a niche market.  Currently, `widely deployable' is somewhat
synonymous with `runnable on the IA-32 architecture or its
derivatives\footnote{here we expressly refer to the AMD64 (or EMT64)
64-bit extensions to the IA-32 ISA},' owing to the fact that Intel and
AMD processors are widely deployed on everything from desktops to
high-performance computing clusters.  While Intel's IA-64 architecture
offers high performance, usage of these processors is currently
limited to the high-performance and scientific arenas.  Another aspect
of the portability issue is software-related: any on-demand computing
system with the ability to run existing scientific and
high-performance code with acceptable performance will certainly see a
greater rate of adoption than those which require massive rewrites.

Xen succeeds in both respects: The VMM runs on the IA-32 ISA and its
64 bit extensions architectures.  While the OS in such a
paravirtualized architecture requires modification in order to work
with Xen, most applications can run unmodified.  As noted above, the
porting effort required to make an operating system `Xen-compatible'
can be managed by a relatively small group, and ports to several
UNIX-like operating systems.  A port of the Windows operating system
was only partially completed as documented in~\cite{dragovic03xen}; 
furthermore that
port was specific to an older version of Xen.  The lack of support for
Windows is not as great a concern in the context of on-demand computing
as in other aspects of software development; first, many
high-performance computing clusters run variants of Linux on their
nodes, for example (perhaps come up with a figure for how many of the
top 500 clusters run Linux).  Secondly, the 3.0 release of Xen will
feature support for Intel's `Vanderpool' hardware virtualization
support, which will allow Xen to virtualize unmodified operating
systems.  Support for high-speed interconnects may be an issue in
virtualized systems.  In Xen, virtual devices for unprivileged virtual
machines (or domains in Xen terminology) are all backed by a physical
device running in a privileged domain.  No support currently exists
for directly accessing Myrinet cards from within an unprivileged
domain, though such an addition to Xen is feasible, as discussed
in~\cite{xenmyrinet}. However, most devices which have Linux drivers can be
recompiled and run unmodified under a privileged Xen domain.  Using
the IP-over-Myrinet driver available for Linux (among other operating
systems), a virtual network device should be able to use Myrinet
hardware as a backend, and thus take advantage of the high throughput
available on these devices, if not the low latency.  Testing is
necessary to see whether I/O is bounded due to high CPU utilization as
observed in~\cite{menon05xen} when using special-purpose high-speed
interconnects.
Denali, like Xen, runs on the IA-32 architecture, but until recently
the paravirtualized Denali architecture disallowed the porting of
general-purpose operating systems.  More recently, the addition of a
virtual MMU to the paravirtualized Denali architecture has
changed this situation, and a port of NetBSD exists for the Denali
VMM.  This support is a more recent development than the
existence of a NetBSD port for Xen, and no port of Linux or any other
UNIX work-alike currently exists for Denali.

VMWare's virtualization products are all available for IA-32 machines.
Along with Xen, this IA-32 support extends (at least for the ESX
Server edition) to multiprocessor machines.  Denali is noted
in~\cite{whitaker02scale}  
as not supporting multiprocessors at that time of writing, though the
design of the VMM is said to allow for such an extension.

Linux-VServer modifies the guest OS to support enhanced security
isolation properties, and as such, has little or no machine-level
dependencies in the Linux kernel.  Ports of Linux-VServer work on
Linux on a variety of architectures, including IA-32, IA-64,
MIPS32/64, HPPA, and PPC.  User-Mode Linux is apparently less
portable.  While an IA-32 version has been stable for some time, a PPC
port in working condition is unavailable as of early 2005, as noted
in~\cite{dike05umlppc}.

The other systems reviewed in this paper are directed towards
execution on special architectures: Disco is targeted at the Stanford
FLASH Machine, and its successor Cellular Disco had a working
implementation on the SGI Origin 2000.
Table~\ref{port_comparison_table} contains a summary of this
section's results.  Note that exact figures for code size were not
mentioned in any of the works on virtualization cited, so the below are
estimations based on further investigation.

\begin{table}
\caption{Portability Comparison}
\begin{center}
\begin{tabular}{| p{2in} | p{2in} | p{2in}|}\hline\label{port_comparison_table}
\textbf{System}	& \textbf{Porting Required} & \textbf{Ports Available}\\
\hline
VMWare ESX Server& No porting required & n/a (any x86 guest OS)\\
\hline
Denali & OS porting required & NetBSD\\
\hline
Xen    & OS porting required  & FreeBSD, Linux, NetBSD\\
\hline
VServer & Linux-only & Linux\\
\hline
User-Mode Linux & Linux on x86 only & Linux\\
\hline
Disco & guest OS modification required& IRIX on FLASH machine simulator only\\
\hline
Cellular Disco &guest OS modification required&IRIX on SGI Origin 2000, Stanford FLASH Machine\\
\hline
\end{tabular}
\end{center}
\end{table}

\subsection{Multiplicity}\label{perf-multiplicity}
One advantage offered by VMMs in the context of on-demand
computing is the combination of efficient utilization and granularity of
sharing.  A common method of resource allocation in high-performance
computing clusters is on a node-by-node basis.  That is, batch job
submissions specify a number of nodes on which to run, with one node
as a minimum.  In addition, no more than one job is generally allowed
to run on one node.  In batch clusters, the amount charged for
resource usage is proportional to the amount of time spent running a
job.  The `one job-one node' rule is used to enforce fairness to
users, as multi-user operating systems do not provide sufficient
resource isolation to prevent multiple jobs on a node from interfering
performance-wise with one another.  Batch schedulers on HPC clusters
must therefore fairly allocate resources to incoming requests with the
above constraints in mind.  Efficient scheduling of resources in this
manner is a hard problem, and can run into inefficiencies in certain
situations.  The Faucets scheduler~\cite{kale04faucets}, which
is built on top of the Charm distributed object
system~\cite{kale93charm}, provides
an interesting solution to the problem of resource utilization in a
cluster context through the use of adaptive jobs: each distributed is
broken into a large number of event-driven objects called
\textit{chares}, and chares are mapped to
physical processors, frequently in a many-to-one relationship.
Through the use of the chare migration functionality built into Charm,
Faucets can resize jobs according to changing utilization across the
nodes of a cluster.  In the interest of performance, the single-node
single-job rule is still enforced at any one time.

The above description of Faucets and Charm provides an example where
virtualization as discussed in this paper can benefit users of HPC by
allowing conventional distributed computations to be made more
adaptable through the use of indirection.  The example above in the
description of Faucets could, through a virtualized resource layer, be
generalized to a wider array of applications.  A number of desirable
features of VMM systems for use in on-demand computing can be derived
from the above discussion:

\begin{itemize}
\item{Isolation}: how effectively are all resources, including CPU,
network, accessible from a virtual machine isolated from the effects
of other virtual machines running same hardware?

\item{Efficiency}: what is the overhead of the VMM with regards to
running multiple virtual machines on the same hardware?  An execution
slowdown proportional to the number of VM instances running is
desirable, but some systems can do better as described below.
\end{itemize}

Linux-VServer builds isolation mechanisms into a Linux kernel.  In
contrast to Xen, memory pages of various systems are not independent,
so the potential exists for sharing of pages through shared libraries,
etc. between sandboxed programs on a machine.  Linux-VServer is
intended for security-related isolation of server programs on the
Linux platform, and extends the file system isolation of the chroot
system call by providing similar isolation for processes and network
interfaces.  Potential sharing of memory increases the efficiency of
Linux-VServer with regards to memory sharing.  Additional memory
sharing and context switching efficiency is realized because all
processes in a system partitioned with Linux-VServer make requests to
the same kernel.  In a Xen system, for example, $n$ copies of the kernel
must exist for $n$ virtual machines.  In theory, context switching
between virtual machines is a more expensive operation than a standard
process context switch, though the Xen researchers estimate that even
with this overhead, a system running 128 virtual machines would only
experience 7\% overhead from the VMM.  With regards to isolation, it should
be noted that Linux-VServer is a project geared only towards security
isolation of user-level processes.  The Linux kernel is not extended
in any way with regards to resource isolation, and so only the
`standard' guarantees are provided in this case.

Xen as noted above is less efficient with regards to memory than
OS-level isolation schemes such as found in Linux-VServer.  However,
Xen is noted even in the VServer literature as being one of the first
contemporary VMM systems to provide resource isolation guarantees.  In
early versions of Xen, these isolation mechanisms were
geared at fair sharing between domains, though differentiated service
was noted as a future work, and should be achievable within the Xen
I/O framework.  Sharing is enforced for CPU utilization via a biased
virtual time-based domain scheduler.  Network and disk requests are
brokered by Xen.  The Xen I/O scheduler services requests from domains
in a round robin fashion.  In the case of disk I/O requests, the
driver domain can
reorder requests to take advantage of disk geometry, and domains can
batch requests to further reduce overhead.  The end result is that
under testing, up to 16 domains running the same resource-intensive
benchmark experience fair sharing between domains.
Domains also benefit from very low overhead from the VMM under these
conditions.  Denali's stated goals are similar to the results 
achieved in Xen, "approximate resource fairness across services."
However, the implementation in~\cite{whitaker02scale} is much more immature.
Virtual disk devices are not available to Denali services, so a
comparison of disk performance is not given.  No testing is done
in~\cite{whitaker02scale} or~\cite{whitaker05rethinking}
with regards to measuring the level of fairness actually
achieved between virtual machines.

Disco~\cite{bugnion97disco} implements its own I/O devices, in a manner similar to
Denali.  As in Denali, an idealized device interface is provided to
make adding devices to guest OS instances a simpler matter.  Denali,
by implementing its own I/O devices, is required to interpose on all
I/O requests, and takes advantage of this to encourage transparent
page-sharing among virtual machines: when remapping VM I/O requests to
hardware addresses, the VMM can detect previously cached disk requests
from other machines, and return a reference to the cached page in case
on such a cache hit.  In conjunction with a virtual networking system
which similarly takes advantage of cached network accesses by other
machines (in particular, network requests for disk blocks via NFS),
Disco maintains a global buffer cache.  The end behavior is that
interacting VM instances end up sharing memory pages, and thus utilize
global memory resources more effectively.  With regards to scheduling,
only the existence of a `simple, time-sharing' scheduler is noted,
 and no comment is made on the resource isolation properties of the VMM.

It should be noted that simply multiplexing resources on individual
machines is not sufficient to provide backwards-compatibility for
parallel computations even if such resource partitioning can be done
which perfect efficiency.  In fact, virtualizing traditional parallel
computations naively can potentially lead to severe performance loss. 
\cite{koenig05using} shows that for parallel codes with
tightly-coupled communication 
patterns, heterogeneous network interconnects can have a negative
effect on overall throughput, and proposes restructuring computations
to use a latency-tolerant parallel computing model (namely the Faucets
scheduler on top of Charm) to mitigate these effects.  Scheduling VM
instances on multiple physical machines independently can cause
unpredictable latency on network links between virtual machines, and
can thus cause similar performance degradation.  Cellular
Disco~\cite{govil00cellular}, 
which can virtualize SMP systems on the NUMA SGI Origin machine,
gang-schedules the virtual CPUs of multiprocessor virtual machines to
avoid latency-dependent performance degradation, but only deals with
the case of multiprocessor scheduling of a individual machines.
VMWare ESX Server currently provides `Virtual SMP' functionality to
SMP virtual machines running on (physical) multiprocessor systems, and
must presumably gang-schedule virtual CPUs in a similar manner to that
of Cellular Disco. The SMP VM support in both machines is geared
towards preventing performance degradation caused if spin locks held by
an unscheduled VM CPU block computations in other VM threads. Neither
of the above systems co-schedule computations on different machines.
None of the systems surveyed in this paper are known to provide such
inter-machine gang-scheduling, and thus a latency-tolerant system such
as Charm must presumably be used on top on a VMM-based on-demand
computing system in order to allow computations on an over-utilized set
of resources to proceed efficiently.
In Table~\ref{isolation_comparison_table}, results
pertaining to the degree of multiplicity supported on each of the
systems reviewed are presented.  While we cannot give exact figures as
to the number of VMs supported on any of these systems (except for ESX
Server, which is limited to 64), we can categorize the systems
reviewed in terms of desirable features.

\begin{table}
\caption{Degree of Multiplicity}
\begin{center}
\begin{tabular}{|p{1.5in}||p{1.5in}|p{1.5in}|p{1.5in}|l|}
\hline\label{isolation_comparison_table}
\textbf{System}	&\multicolumn{3}{c}{\textbf{Isolation Provided}}&\\
\hline
~ & \textbf{Level of Isolation}& \textbf{Resource} & \textbf{Security} &\\ 
\hline
VMWare ESX Server & Machine & Yes & Yes&\\
\hline
Denali         & Machine         & Yes & Yes &\\
               & Paravirtualized &     &     &\\
\hline
Xen            & Machine         & Yes & Yes&\\
               & Paravirtualized &     &    &\\
\hline
VServer        & OS-level   & No & Yes&\\
\hline
User-Mode Linux& OS-level   & No & Yes(in SKAS mode)&\\
\hline
Disco          & Machine         & CPU isolation provided. & Yes&\\
               & Paravirtualized & Unknown if network and disks are as
well&     &\\
\hline
Cellular Disco & Machine         & CPU isolation provided. & Yes&\\
               & Paravirtualized & Unknown if network and disks are as
well&     &\\
\hline
\end{tabular}
\end{center}
\end{table}

\subsection{Trust}
The discussion of security in this paper avoids a detailed threat
modeling and vulnerability analysis of a secure On-Demand Computing
infrastructure, and we defer such analysis to a later work.  Several
virtual machine-based systems have been used to enhance the security
of functions such as logging and auditing by building such
functionality into the VMM
layer~\cite{seiden90auditing,dunlap02revirt}.
However, these
systems are either geared primarily towards security and not
performance, or do not contain a detailed security analysis.  Other
systems such as Xen and Denali note security as a benefit of their
respective VMM architectures, but again are not focused

Most of the works cited in this paper do not perform extensive
security analyses of the architectures they cover.  As such, it is
difficult to cover in detail the security-related benefits of the
system and OS-level VMM systems herein.  Thus the survey of these
systems from a security standpoint in rather general.  

Access control and isolation of resources are essential
responsibilities of any modern multi-user operating system.  Security
in modern operating systems depends on the ability of the OS kernel to
reliably provide protection to a running process from other, possibly
misbehaving processes.  On the one hand, this involves preventing
processes from writing into the address space of other processes, a
feature supported by all modern ISAs through the use of paging and/or
segmentation.  In addition, the OS should protect the machine from
harm through device malfunctions caused through malicious process
behavior.  The latter of the above criteria should only be provided to
the extent that hardware protection allows it, of course.  In general,
the security provided by an operating system kernel, within the
constraints of hardware protection, relies upon the following:
\begin{itemize}
\item The interfaces to resources provided by the operating system
should be 
consistently secure, in the sense that user-space processes should not
be able to exploit `legal' sequences of system calls to obtain
unexpectedly high privileges.  The term `legal' is intended to imply
that the consistency of OS design, and not any security bugs, are
being considered in this context.  The above will be referred to as
the \textit{architected} security properties of the operating system.

\item The implementation of the operating system services provided to
user-space processes should be free of bugs.  Properties dealing with
the quality of implementation provided by the operating system with
regards to the bugs in the implementation will be referred to as the
\textit{implemented} security properties of the kernel.
\end{itemize}

The actual degree of security provided by the OS depends on both of
the above.  The degree of architected security depends on the
consistency in the design of the OS system call interface, and the
degree to which system call semantics are well defined.  The
architected security of an operating system depends on the degree to
which the operating system faithfully implements the security-related
semantics of the provided OS interface.  In practice, the design of
the OS interface presents a simpler problem: the hardware features of
modern processors let the OS define a very small interface by which
user-space processes may access the services of the OS, making the
security interface design a manageable problem.  The implementation of
these services presents a more difficult problem:  faithful
implementation of the defined security policies depends on internal OS
checks with regards to the resources being requested from the OS, and
with regards to the privileges available to the requesting entity.  As
many of these checks are done purely in software, no hardware
assistance is available, and thus security depends on system code
being bug-free.  Thus the OS code in general-purpose operating systems
is treated as a \textit{trusted layer}.  This assumption is
generally difficult to verify for modern operating systems, which,
including device drivers, can include millions of lines of privileged
code.  Software engineering techniques can alleviate the
complexity of verifying system software integrity somewhat: strict
internal interfaces can be designed for given subsystems of an OS
kernel.  However, this approach suffers from a number of problems:
first, conformance to these interfaces can be difficult, especially if
internal interfaces can change often, as is the case in the 2.6 Linux
kernel.  In addition, conformance to interfaces
is voluntary: in the absence of hardware protection, no guarantees can
be made that a provided interface is being used.  The first problem
can be dealt with by deprecating unused interfaces.  Out of date code
will then generate warnings or refuse to compile at all, letting the
compiler act as a security check of sorts.  The problem of protection
in internal kernel services has been dealt with in a number of ways.
One of the better-known approaches is a microkernel:
the kernel is restructured as a minimal entity which is only
responsible for providing a very basic set of services, such as access
to hardware and a message routing framework, and all higher-level OS
services, such as memory management, scheduling policies, etc., are
delegated to privileged user-space tasks which can call on the
services of the kernel.  This loosely coupled design keeps the trusted
interface very small, allowing for a simpler and thus more verifiable
secure trusted system software base.  In practice, performance is
usually an issue, as every request from a non-privileged process
results in two context switches. 

Virtual machine monitors provide a potentially attractive solution to
the problem of security vs. performance in the context of trusted
system software layers.  The abstractions provided by many of the
systems discussed in this paper are either identical to or very
similar to those provided by real hardware.  The low level of service
provided can reduce implementation complexity considerably: the Disco
VMM was made up of 13,000 lines of code, and the more fully-functional
Xen VMM\footnote{the original, 13,000 line version of Disco was tested
on a simulator; Cellular Disco, which ran on real hardware, weighed in
at 50,000 lines of code, and piggybacked on top on an existing OS (IRIX
6.4) for device access.} is of similar size.  Moreover, VMMs can offer
near native performance for guest VM code, while already performing
checks on resource allocation.  The VAX Security kernel
and IBM's SHype architecture are existing systems which use
virtual machine monitors to provide a trusted resource management
layer to guest programs, though neither of them consider
performance issues.

The method by which resource control can be obtained with low overhead
is hinted at in the Xen paper and more explicitly stated in the
Exokernel work:
instead of performing resource usage checks at every use of a given
resource, such checks are limited to resource \textit{binding} time:
a VM is required to register resource usage only once with the VMM, at
which point the manner is which the resource is used is largely
delegated to the VM.  In both cases, while use is left up to the VM
(or libOS in the case of Exokernel), the VMM remains in control; the
Exokernel design specifies a \textit{revocation protocol} for removing
a VM's access rights to a given resource; any VMs which do not
cooperate can be forcibly terminated by the VM.  In Xen, inter-VM
communication comes through shared memory mappings managed by the VMM;
access to regions not authorized by the VMM can be trapped by hardware
and recognized by Xen, which terminates the guilty VM.

Linux-VServer provides a study of a case where virtualization is done
at other than the (idealized) machine layer, in this case at the OS
level.  The interface at which security must be enforced in this case
is still small; the Linux system call interface, for example, is narrow and
well-defined.  However, the required security checks for user-space
processes must be integrated into the kernel proper in this case, so
many of the problems mentioned above with respect to verifying the
integrity of the trusted system layer remain in the case of VServer.

In summary, system and OS level virtualization alleviate the
performance problems of microkernel systems in different ways: in a
traditional virtualized architecture, only the the binding of
resources to a VM is managed by the VMM, with all usage totally
unmediated by the monitor.  This provides low overhead, but at the
cost of granularity of control.  This is noted in SHype, which
appealed to the use of a finer-grained resource control mechanism
inside each guest instance to deal with the problem.  OS-level
virtualization provides access control at exactly the level of
granularity of the OS system call layer.  Overhead is less than in a
microkernel system because all access checks are done in the kernel
context, but bugs in the operating system are more likely to affect
the security guarantees provided in this case.

Due to the inexact nature of the analysis in this section, we refrain
from attempting to quantify the level of security in any of the
systems covered in this section.  We do present some figures
on code size and maturity which may pertain to security auditing of
these systems in Table~\ref{sec_comparison_table}.

\begin{table}
\caption{Trust Comparison}
\begin{center}
\begin{tabular}{| p{2in} | p{2in} | p{2in} |}\hline\label{sec_comparison_table}
\textbf{System}	& \textbf{Size of Code Base} & \textbf{Maturity}\\
\hline
VMWare ESX Server & unknown & Commercial, in production\\
\hline
Denali &\~20,000 lines & n/a\\
\hline
Xen & \~30,000 lines &  Production\\
\hline
VServer & 2,400 line patch + security tools & In production\\
\hline
User-Mode Linux & 45k line patch + 630line patch for SKAS mode& In production\\
\hline
Disco &13k lines & Experimental\\
\hline
Cellular Disco & 50k lines & Experimental\\
\hline
\end{tabular}
\end{center}
\end{table}

\section{Comparison \& Conclusion}
In this paper we have reviewed various schemes for virtualization and
process isolation on a number of system architectures.  Next we
present a brief summary of the considerations we made in choosing a
final candidate as the basis for our On-Demand Secure On-Demand
Computing project.

\subsection{Comparison}
From the brief review above, it can be seen that some of the systems
below are clearly not appropriate for the purposes of building a
on-demand computing infrastructure building on commodity hardware and
software.
\begin{itemize}
\item \textbf{Disco} - The original Disco implementation is
interesting from a research perspective.  However, the implementation
described in~\cite{bugnion97disco} was only tested on a simulator for the
Stanford FLASH architecture (which was the basis for the SGI Origin
machine).  Minimal testing was done on real hardware, and that only on
a single CPU machine.  As the targeted architecture, a scalable
multiprocessor, also does not fall within the scope of commodity
hardware, we do not consider Disco further.
\item \textbf{Cellular Disco} - The implementation
in~\cite{govil00cellular} is more promising than that of Disco - testing
was done on real hardware.  However, the target architecture of this
system, the SGI Origin 2000, is also not under the umbrella of
commodity hardware.  Moreover, like Disco, the guest instances of
SGI's IRIX required modifications not released to the public.  As is
the case with Disco, therefore, we do not consider Cellular Disco
further.
\item \textbf{User-Mode Linux} - UML is mature and readily available
for download.  Moreover, any standard Linux distribution can serve as
the basis for a guest OS instance once the appropriate guest kernel
modifications are made.  The performance tests
in~\cite{dragovic03xen} show that
while CPU intensive code can run largely at full speed, the execution
of code which makes frequent system calls (as in the case of I/O
intensive programs) can be slowed considerably\footnote{This assumes
the use of Separate Kernel Address Space mode; running a UML guest on
an unmodified kernel results in considerably reduced performance}.
\item \textbf{Denali} - The scope of the Denali project has changed
from its original incarnation:  While the original version targeted
specialized libOS guests, the addition of a virtual MMU allowed the
development of a guest port of NetBSD.  As one of the goals of Denali
was to allow the efficient execution of a large number of guests, it
can be assumed that execution of guests occurs at a significant
fraction of native speed.  However, the performance figures
in~\cite{whitaker02scale} focus on the scalability of the VMM as the
number of 
guests increases, and not on a direct comparison of hosted and native
performance.  Another disadvantage is hardware support: as Denali runs
directly on x86 hardware, it must directly interface with hardware and
I/O devices.  Thus, it is likely that supporting a wide range of
hardware would be more difficult with Denali than with a system such
as Xen, which defers direct hardware access to privileged 
\textit{I/O domains}.  As Xen has wider industry support, and
will be integrated into the mainline Linux kernel, we favor that
system over Denali in further discussion.
\end{itemize}

The remaining systems to be considered are VMWare ESX Server,
Linux-VServer, and Xen.  VServer is widely used to provide security
isolation for hosting services, and is in fact used in the PlanetLab
system~\cite{peterson03planetlab}, to provide isolated \textit{slivers}
of computing resources for use by planetary-scale distributed computing
services\footnote{the Emulab system, which is intended for testing
distributed systems in as in PlanetLab, currently provides the
option of using the FreeBSD jail mechanism to provide a degree of
virtualization on hosts.  As Linux-VServer is very similar in concept,
we don't discuss jail further.}.  PlanetLab's goal seems at least
partially in line with our 
own, and so it would seem that VServer could also be adapted for our
own usage.  However, the VServer only provides security isolation, and
it is through the use of the SILK~\cite{bavier02silk} module that resource
isolation is achieved in PlanetLab.  The SILK module provides CPU and
network isolation to processes on a Linux system, but the disk
resource is not similarly isolated.  Also, applications require some
modification to receive the resource isolation benefits of SILK.

VMWare ESX Server and Xen are both attractive candidates for use in
our On-Demand Secure Cluster Computing project. In the end Xen was
selected as the basis for future work.  While ESX 
server is noted in the Xen work as having better
performance than the hosted VMWare editions, the same notes that
Xen still outperforms the higher-end VMWare product.
In addition, the lack of performance data
for ESX server (due to licensing restrictions on benchmarking) made an
independent performance analysis of ESX Server impossible.  With
regards to price, we can state that VMWare ESX Server comes with a
significantly higher up-front cost than Xen, which is open-source and
free.  For us this cost was not justifiable, especially given that the
OS modification requirement for guests under Xen was not a problem in
practice: Stable ports exist for both the 2.4 and 2.6 Linux kernel
versions, as well as for a number of open-source Unix variants.  
%
%
\bibliographystyle{plain}
\small
\bibliography{Survey}

\end{document}